\documentclass[useAMS,usenatbib]{mn2e}

\usepackage[]{epsfig}

\title[Theoretical Light Curves of Type II-P SNe and Applications to Cosmology]
{Theoretical Light Curves of Type II-P SNe and Applications to Cosmology}

\author[A. Chieffi et al.]{A. Chieffi,$^{1,4}$, I. Dom\'\i nguez,$^2$, P. H\"oflich,$^3$, M. Limongi,$^4$, O. Straniero,$^5$\\
$^1$ Istituto di Astrofisica Spaziale e Fisica Cosmica (CNR), Via Fosso del Cavaliere, I-00133 Rome, Italy, achieffi@rm.iasf.cnr.it\\
$^2$ Dept. F\'\i sica Te\'orica y del Cosmos, Universidad de Granada, E-18071 Granada, Spain, inma@ugr.es\\
$^3$ Dept. of Astronomy, University of Texas, Austin, TX78681, USA, pah@astro.as.utexas.edu\\
$^4$ INAF-Osservatorio Astronomico di Roma, Via Frascati 33, I-00040 Monteporzio Catone, Italy, marco@mporzio.astro.it\\ 
$^5$ INAF-Osservatorio Astronomico di Teramo, I-64100 Teramo, Italy, straniero@te.astro.it}
\date{Released 2002 Xxxxx XX}

\pagerange{\pageref{firstpage}--\pageref{lastpage}} \pubyear{2002}

\def\LaTeX{L\kern-.36em\raise.3ex\hbox{a}\kern-.15em
    T\kern-.1667em\lower.7ex\hbox{E}\kern-.125emX}

\begin{document}

\label{firstpage}

\maketitle

\begin{abstract} 
 Based on an extensive grid of stellar models between 13 and  25
 $\rm M_\odot$ and a wide range of metallicities, we have studied  the 
 light curves of core  collapse  supernovae, their application to cosmology 
 and evolutionary effects with redshift. The direct
link between the hydro and radiation transport allows to calculate monochromatic
light curves. 

With decreasing metallicity Z and  increasing mass, progenitors tend to  explode
 as compact Blue Supergiants (BSG)  and produce sub-luminous supernovae that  are
 about $1.5^m$ dimmer compared to {\it normal} SNe II  with Red Supergiant (RSG)
 progenitors. Progenitors with small masses tend to explode as RSGs even at  low
 Z. The consequences are obvious  for probing the chemical evolution,  namely, a
 strong bias when using the statistics of  core collapse supernovae to probe the
 history of star formation.

 Our study is limited  in scope with respect to the explosion energies and the production of 
radioactive Ni.  Within the class of {\it extreme SNe~II-P} supernovae,  the light curves
are rather insensitive with respect to the progenitor mass and explosion energy compared
to analytic models which are based on parameterized stellar structures.
 We expect a wider range of brightness due to variations in $^{56}$Ni because radioactive energy is
a main source of luminosity. However, the overall insensitivity of LCs
may allow their use as  quasi-standard candles for distance determination. 
\end{abstract}
 

\section{Introduction}

  Core  collapse supernovae, are thought to be  the final
 results of stellar evolution  for stars with main  sequence masses $\ga$  10
   $\rm  M_\odot$ with  evolutionary time  scales short  compared to  the age  of the
 universe even at high z (Tammann, 1982; Maza \& Van den Bergh, 1976; Woosley \&
 Weaver 1986). These  objects will occur  soon after the  initial star formation
 period and, therefore, can  be used to probe  the structure of the  universe at
 high {\em z}. These very distant supernovae are all expected to be some variety
 of core collapse supernovae.  E.g. at $z \sim  5$ to $10$, galaxies are expected  to be
 small and dim and core collapse supernovae may be the brightest objects in  the
 Universe (Miralda-Escud\'e \& Rees 1997).

The light  curves and  spectra depend  sensitively on  the initial stellar mass,
 metallicity,  mass  loss and  explosion  energy. They   show  a wide  range  of
 brightness,  up  to  6  magnitudes,  and  properties  of  their  light   curves
 (Filippenko, 2000;  Patat et  al. 1993  \& 1994;  Young \&  Branch, 1989)  that
 prevents their use as standard-candles. However, our knowledge of the event  is
 improving and it may be possible to derive the absolute magnitude in a  similar
 way as Type Ia if appropriate empirical correlations can be identified. 

There is a general agreement that the  explosion of a massive star is caused  by
 the collapse  of its  central parts  into a  neutron star  or a black hole. The
 mechanism of  the energy  deposition into  the envelope  is still debated.  The
 process likely involves the bounce and the formation of the prompt shock  (e.g.
 Van Riper  1978, Hillebrandt  1982), radiation  of the  energy in  the form  of
 neutrinos (e.g. Bowers \& Wilson 1982),  the interaction of the neutrinos  with
 the material of the  envelope and  various types  of convective motions (  e.g.
 Herant et  al. 1994,  Burrows et  al. 1995,  M\"uller \&  Janka 1997,  Janka \&
 M\"uller 1996), rotation  (e.g. LeBlanc \& Wilson 1970, Saenz \& Shapiro  1981,
 M\"onchmeyer et al. 1991),  and  magnetic fields (e.g. LeBlanc \&  Wilson 1970,
 Bisnovati-Kogan 1971).

The analysis of core collapse supernovae and their use for cosmology is further complicated due  to
 the mounting evidence that the  explosions of massive stars (core  collapse SN)
 are  highly  aspherical  and, consequently,  the  brightness  depends on  their
 orientation  to the  observer (e.g.  H\"oflich 1991b).  (1) The  spectra  (e.g.
 SN87A, SN93J, SN94I, SN99em) are significantly polarized indicating  asymmetric
 envelopes (M\'endez  et al.  1988; H\"oflich  1991b; Jeffrey  1991; Wang et al.
 1996; Wang, Wheeler  \& H\"oflich 1999).  The degree of  polarization  tends to
 vary  inversely with  the mass  of the  hydrogen envelope,  being maximum   for
 SNe~Ib/c events with no hydrogen (Wang et al. 2000, Leonard et al. 2002). For supernovae, with a good
 time and wavelength coverage,   the orientation of the polarization vector tends
 to stay constant both in time  and in the wavelength. This suggests  that there
 is  a   global  symmetry  axis  in  the  ejecta  (Leonard  et  al.  2001).  (2)
 Observations of SN~1987A  showed that radioactive  material was brought  to the
 hydrogen rich layers of the  ejecta very quickly during the  explosion (Tueller
 et al.  1991). (3)  The remnant  of the  Cas~A supernova  shows rapidly  moving
 oxygen-rich matter outside the nominal boundary of the remnant              and
 evidence for two oppositely directed jets of high-velocity material (Fesen   \&
 Gunderson 1997). (4) Recent X-ray observations with the CHANDRA satellite  have
 shown an  unusual distribution  of iron  and silicon  group elements with large
 scale asymmetry in Cas~A (Huges et al. 2000). (5) After the explosion,  neutron
 stars are observed with high velocities up to 1000  km/s (Strom et al. 1995).

Due to the difficulty  of modeling core collapse  from first principles, a  very
 different line of attack on the explosion problem has been used extensively and
 proved to be successful in  understanding the supernova problem,  SN1987A in
 particular (Arnett et al. 1990,  Hillebrand \& H\"oflich 1991). The  difference
 of characteristic time scales of the  core (a second or less) and  the envelope
 (hours to days)  allows one to  divide the explosion  problem into two  largely
 independent parts  - the  core collapse  and the  ejection of  the envelope. By
 assuming the characteristics of the energy deposition into the envelope  during
 the core collapse, the response of the envelope can be calculated. Thus one can
 study   the   observational   consequences   of   the   explosion   and  deduce
 characteristics  of  the  core  collapse  and  the  progenitor  structure. This
 approach has been  extensively applied in  the framework of  the 1D spherically
 symmetric  formulation. The  major factors  influencing the  outcome have  been
 found to be  the explosion energy  and the progenitor  structure. Recently, the
 same approach has been applied  in multi-dimensions to investigate the  effects
 of asymmetric explosions (e.g. H\"oflich et al. 2001). First results show  that
 both  asymmetric  density   structures  and  excitations   are  keys  for   our
 understanding of the global asymmetries in core collapse supernovae.

Despite these  problems, methods  have been  developed to  use these objects for
 distance  determinations,  namely,  SNe~II  that  have  retained  their  H-rich
 envelope. SNe~II represent an important complement to SNe~Ia as a technique  to
 measure cosmological distances. In  principle, the distance to  each individual
 SNe~II  can  be  measured  by  the  {\it  Baade-Wesselink}  or  {\it  Expanding
 Photosphere} method in  which the ratio  of the observed  apparent flux to  the
 absolute flux gives the solid angle, and the velocity and elapsed time give the
 radius (e.g. Shaviv  et al. 1985,  H\"oflich et al.  1986, Branch et  al. 1988,
 H\"oflich 1988, 1991, Schmidt et al. 1994, Leonard et al. 2002). 
 
Recently a sub-group of SNe~II,  SNe~II-P,  has been proposed as good  quasi-standard
 candles (H\"oflich et al. 2000,  Hamuy and Pinto 2002), decreasing  the scatter
 in the Hubble diagram. The plateau  in the SN~II-P light curve, that  is caused
 by the recombination of H, is indicative of a massive H envelope. Thanks to the
 plateau the identification does not require any spectra and it is not necessary
 to follow the event  to dimmer phases. We  identify a subclass among  SNe II-P,
 {\it extreme} SNe II-P, characterized by a {\it long} plateau, with a  duration
 exceeding  50-60  days. Notice  that  the characteristic  time  of the  plateau
 observed by the telescopes  is increased by a  factor (1+{\em z}) (ie.  at {\em
 z}=3, the {\it bright} observed plateau lasts several months). 
 
Although, their statistic is still  very incomplete, {\it extreme} SNe  II-P are
 probably as   frequent as  SNe Ia  in the  nearby universe  (Cappellaro et  al.
 1999). In  the years  1998 and  1999 around  a 10$\%$  of all  SNe II  could be
 classified as  {\it extreme}  SN~II-P. However,  this is  expected to change at
 {\it intermediate} redshifts ($z\approx$ 3-5) when the star formation rate was
 higher (Kravtso \& Yepes, 2000). At  the other hand, SNe~II maybe the  only SNe
 at $z\ge$3-5.  Moreover, at  $z\ge$10, they  may be  the only luminous objects,
 out-shining the host proto-galaxy by a  factor of 100 during more than  2 years
 (Marri \&  Ferrara 1998).  Cold Dark  Matter (CDM)  models, taken  into account
 gravitational magnification, predict in 100 fields of the Next Generation Space
 Telescope around 850  and 3500 SNII/year  (for Standard and  Lambda CDM models,
 respectively) up to a redshift of $z$=15 (Marri, Ferrara \& Pozzetti, 2000). 
 
One key problem to be addressed  will be to determine whether these  events will
 look like local supernovae that explode as red supergiants, or whether, because
 of their low  metallicity they will  all resemble SN  1987A that exploded  as a
 blue supergiant (see also Brunish \& Truran 1982). Prediction of the properties
 of these  early supernovae  now will  set the  stage for  the work  of upcoming
 observational facilities, like NGST, SNAP and others. 
 
Very few  studies of  light curves  of SNe~II-P  have been  published up to now.
 Litvinova \&  Nad\"ezhin (1983,1985)  presented a  grid for  a range of envelope
 masses between 1 and 16 $\rm M_\odot$ that, some 20 years later, still presents
 the  most  comprehensive   study.   The models assumed spherical symmetry and parameterized
density profiles, and 4 basic free parameters: input energy,  the amount of radioactive $Ni$
and the envelope mass and initial radius. The result
of these calculations were theoretical bolometric lightcurves which
were then converted into V-band lightcurves assuming a black body energy distribution.
 Some of these restrictions  have
 been overcome in the analysis of Eastman  et al. (1994) but this study has  been
 restricted to a star with 15 $\rm M_\odot$. Both studies cannot answer the  key
 problems mentioned above.
 In this paper, we link the  pre-SN evolutionary models to SN models. One big advantage
is that the initial  radius of the SN models is no longer an adjustable parameter but
a result of the physical model and a function of the stellar
mass. The other   progress is that the hydro models are linked to a radiative transport code,
 so the broad-band lightcurves are much more accurately computed.
A third  progress provided by this paper is the effect  of metallicity in the resulting 
lightcurves.
 
Following the successful  approach to assume  the characteristics of  the energy
 deposition at  the inner  regions of  the exploding  star, we present explosion
 models and  theoretical light  curves for  a wide  range of  stellar masses and
 metallicities.  The  goal  of  this study  is  to  explore  the sensitivity  of
 different  characteristics  of  the  SN~II-P  light  curves  on  the underlying
 progenitor properties  (initial mass  and metallicity)  and the  kinetic energy
 based on detailed stellar evolution  and LC calculations. Effects due to stellar 
rotation and mass loss  during the  evolution of the  progenitor are not  included.
 Moreover, these calculations do not address the  mechanism of energy deposition, production
of radioactive Ni, chemical
 mixing and departures from sphericity. One of  our main results  is the
 identification of {\it extreme} SN~II-P,  as potential candidates to be  use in
 cosmology. Note that we do not intend  to provide a complete grid of models  to
 be  compared with  specific observations.  In this  work, we  present a   study
 focused on core collapse supernovae  to answer the following questions:  How do
 the light curves of core collapse supernovae depend on the metallicity that  is
 expected to decrease with redshift? Can  we identify a subclass among the  core
 collapse  supernovae  that may  be  used as  quasi-standard  candles, and  what
 accuracy do we expect?  Can this subclass be  identified purely by their  light
 curves, without  a follow-up  that requires  to {\it go}  much fainter than maximum
 light? 

The outline of the paper is as follows: in Section 2 the numerical methods  used
 for the evolution,  explosion and light  curves computations are  described, in
 Section 3 the influence of  the progenitor properties and explosion  parameters
 on the light curves are analyzed, and finally, in Section 4, we highlights  the
 main conclusions and short-comes of our study.

\section{Numerical Methods}
\subsection{Stellar Evolution}

All  the  pre-supernova models   adopted in  this  paper  have been  computed by
 means of the evolutionary code FRANEC, rel 4.2 (for details see \citealt{CS89},
 \citealt{CLS98},  \citealt{LSC00},  \citealt{SCL97}  and \citealt{CL02}).   The
 FRANEC is an hydrostatic evolutionary code  in which both the set of  equations
 describing the physical structure of the star (by assuming spherical  symmetry)
 and the chemical  evolution of the  matter, due to  the nuclear reactions,  are
 fully coupled  and integrated  simultaneously by  means of  a classical  Newton
-Raphson method. The  nuclear network includes  41 isotopes for  the H burning,
 88 isotopes for   the He  burning  and  179 isotopes   for the  more   advanced
 phases. Nuclear  reaction  rates   are taken  from  the   Thielemann's database
 (private communication).  For the   $^{12}{\rm C}(\alpha,\gamma)^{16}{\rm   O}$
 reaction we adopt  the value from  \cite{CFHZ85} which is  close to the   upper
 limit compatible with the presently available measurements  for this rate.  The
 weak  interaction rates   as  a  function   of  the  temperature   and  density
 are  derived   from Fuller, Fowler \& Newman (1980, 1982 and 1985). 

The extension of the convective regions are fixed by means of the  Schwarzschild
 criterion  and no  mechanical overshoot  is allowed.  Induced overshooting  and
 semi-convection  during  core   He  burning  are taken  into    account
 \citep{CCPT85}. In the convective layers the temperature gradient is  evaluated
 by  means of  the mixing  length theory  as described  by \cite{CG68}.  A  time
 dependent mixing scheme is used which is similar to the one firstly  introduced
 by \cite{SE80}. We  have also developed  a specific algorithm  which can handle
 the  evolution  of  those  isotopes  whose  nuclear  burning  lifetime  becomes
 comparable or lower than the mixing timescale (see \citealt{CLS98}).

Electron screenings are taken from \cite{GDGC73} for the weak, intermediate  and
 intermediate-strong regime,  and from  \cite{ITI77} and  \cite{ITID79} for  the
 strong  regime.  The  equation  of   state  (EOS)  is  the  one   described  by
 \cite{Straniero88}   and   updated  by   \cite{SCL97}.   It  includes   Coulomb
 corrections,  quantum-relativistic  effects  on  the  electron  component   and
 electron-positron pairs production. Radiative opacity coefficients are  derived
 from \cite{Kurucz91}, \cite{AF94}, \cite{IRW92} (OPAL) and from the Los  Alamos
 Opacity   Library  (LAOL)   \citep{Huebneretal77}.  A   scaled  solar   mixture
 \citep{Grevesse91}   is  adopted.  Opacity  coefficients  due  to  the  thermal
 conductivity are derived from \cite{Itohetal83}. Energy loss due to photo, pair
 and plasma  neutrinos are  taken into  account following \cite{Munakataetal85},
 \cite{Munakataetal86}.   Bremsstrahlung   neutrinos   are   obtained  following
 \cite{Dicusetal76} as corrected  by \cite{Richardsonetal82}. Energy  losses due
 to  the recombination  processes are  included by  using the  prescriptions  of
 \cite{Beaudetetal67}.

In this work, all the models are evolved at constant mass and no rotation
has been taken into account. The effects of rapid rotation in the evolution of massive stars
 have been recently studied in 1D by Heger, Langer and Woosley (2000) and Meynet and Maeder (2000,2003).
 As discussed in the introduction, the explosion mechanism itself could
be based on rotation but the effect on the hydrodynamics of the H-rich envelope
are expected to be small. However, anisotropies in the
Ni distribution and, thus, the ionization may cause anisotropic luminosities of the order of
 of  10 \% (H\"oflich 1991, H\"oflich et al. 2001).

\subsection{Explosion and Light Curve Models} 
 
The explosions  are calculated  using our  one-dimensional radiation-hydro code,
 including nuclear networks (H\"oflich \& Khokhlov 1996 and references therein).
 This  code  solves the  hydrodynamical  equations explicitly  by  the piecewise
 parabolic method (Collela  \& Woodward 1984)  and includes the  solution of the
 frequency  averaged  radiation  transport  implicitly  via  momentum equations,
 expansion  opacities (see  below), and  detailed equation  of state.  For  high
 densities and  temperatures ($\geq   1 g/cm^3$,  $\geq  10^7$  K), relativistic
 effects are  taken into  account  and  full ionization  is assumed.   For lower
 densities  and  temperatures,   ionization  processes  are  included  under the
 assumption of  local thermodynamical  equilibrium but  relativistic effects are
 neglected.

The explosion is triggered artificially by depositing energy at a mass location 
 near the edge of the iron core.
 After the initial phase of the explosion, i.e. shortly before the shock   front
 reaches the stellar surface, the nuclear reactions are switched off and $\gamma
 $ ray transport is included via  a Monte Carlo scheme.  Both monochromatic  and
 bolometric  light  curves are  calculated  using a  scheme  recently developed,
 tested and  widely applied  to SNe~Ia  (H\"oflich et  al. 1998,  and references
 therein). In order to allow for  a more consistent treatment of scattering,  we
 solve  both  the  (two  lowest)  time-dependent,  frequency  averaged radiation
 momentum equations for the radiation energy and the radiation flux, and a total
 energy  equation.  At each  time  step, we  then  use $T(r)$  to  determine the
 Eddington  factors  and  mean  opacities  by  solving  the  frequency-dependent
 radiation transport equation in the comoving frame and integrate to obtain  the
 frequency-averaged quantities. The averaged  opacities are calculated under the
 assumption of  local thermodynamical  equilibrium. Both  the monochromatic  and
 mean opacities are calculated using the Sobolev approximation. The  scattering,
 photon redistribution  and thermalization terms used in the light curve opacity
 calculation are calibrated  with NLTE calculations  using the formalism  of the
 equivalent-two-level approach (H\"oflich 1995). About one thousand  frequencies
 and between 550 to 700 depth points are used.

\begin{figure}
\includegraphics[width=80mm]{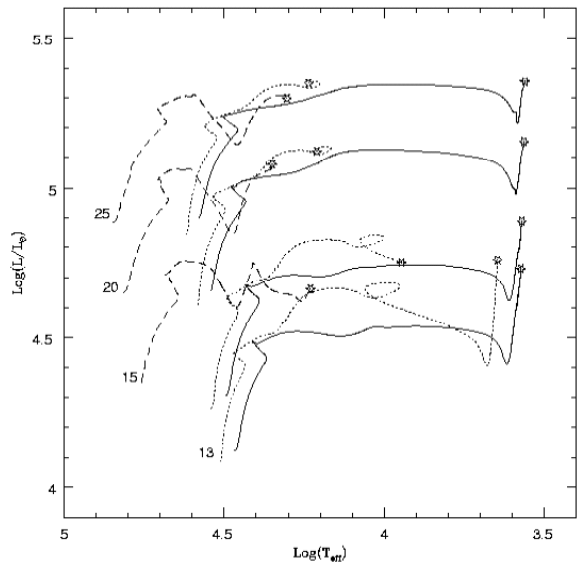}
\caption{Presupernova evolutionary path in the HR diagram of
models 13, 15, 20 and 25 $\rm M_\odot$ having metallicities Z=0.02
(solid), 0.001 (dotted), and 0 (dashed).}
\end{figure}

\section {Results}
 
\subsection{Stellar Models}

The evolution  of selected  models in  the mass  range from  13 $\rm M_\odot$  to 25
 $\rm M_\odot$ and metallicities between Z=0 and Z=0.02 have been computed from  the
 pre-main sequence to the onset of core collapse  without mass loss (see Fig.  1
 and  Table  1).  For  low  metallicities  Z,  models  explode   as  compact BSG
 ($R_\star \leq   100 R_\odot$)   rather than   as extended  RSG  ($500  R_\odot
 \leq    R_\star   \leq   1500   R_\odot$). 
The metallicity plays a major role in determining the radius of the star
 because it affects the opacity that determines the super-adiabatic
  gradient. The larger the opacity, the larger the
 super-adiabatic gradient and hence the larger the radius of the star. This explains why for 
 low metallicities, stars explode as blue supergiants.
 In general, the stellar
 radius depends on the opacity of the envelope; the lower is the metallicity the 
 lower is the opacity of the envelope and, as a consequence, the more compact is the
 structure. 
 We   find   that   all   the   zero
 metallicity models end up as a BSG while all the solar metallicity ones end  up
 as a RSG. At intermediate Z, there  is the general trend that the more  massive
 stars end up as BSG  while the less massive ones  end up as RSG.  The  limiting
 mass depends on Z. However,  the critical metallicity beyond which  a star ends
 as RSG instead  of a BSG  depends sensitively on  many details among  which the
 treatment of convection, the opacities, rotation and  mass loss. 

\begin{figure}
\includegraphics[width=80mm]{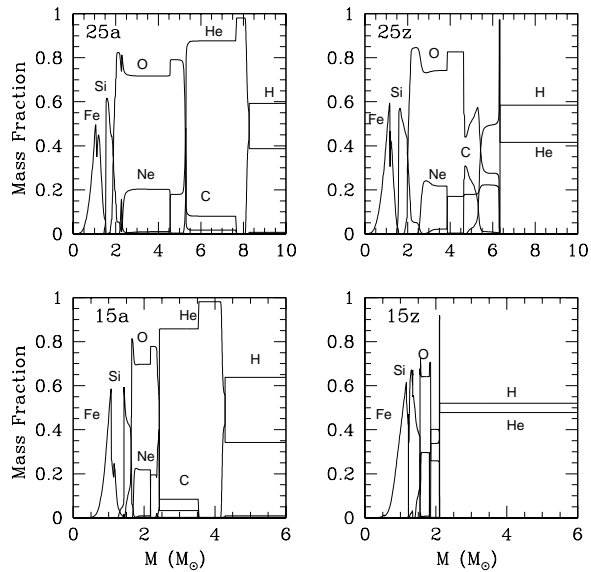}
\caption {Chemical composition of selected models at the onset of 
 core collapse.}
\end{figure}

In  Table  1 we  report  selected properties  of  the presupernova  evolutions,
 namely: the  identification name  of the  models, Mnnx,  nn is  the mass  and x
 refers to metallicity, "a" is for  solar (Z=0.02), "b" for Z=0.001 and  "z" for
 Z=0 (column 1); the mass in solar  units (column 2); the final radius in  solar
 radii (column  3); the  metallicity (column  4); BSG  vs. RSG  (column 5);  the
 hydrogen burning lifetime in years  (column 6); the helium burning  lifetime in
 years (column 7); the residual lifetime following the central He exhaustion  up
 to the  iron core  collapse (column  8); the  final surface  mass fraction of H
 (column 9) and He (column 10).

\begin{table*}
\begin{minipage}{126mm}
\caption{Selected quantities of the some of the stellar  models}
\begin{tabular}{lccccccccc}
\hline
Model &
Mass  &
$\rm R_{\rm fin}~(R_\odot)$  &
Z &
BSG vs. RSG &
$\tau_{\rm H}$(yr)  &
$\tau_{\rm He}$(yr) &
$\tau_{\rm adv}$(yr) &
$\rm H_{\rm sup}$ &
$\rm He_{\rm sup}$ \\ 
\hline
m13a   &  13  & $5.52\cdot10^{2}$ &  $2\cdot10^{-2}$ & RSG & $1.26\cdot10^{7}$ & $1.92\cdot10^{6}$ & $6.57\cdot10^{4}$ & 0.649 & 0.331 \\
m15a   &  15  & $6.71\cdot10^{2}$ &  $2\cdot10^{-2}$ & RSG & $1.08\cdot10^{7}$ & $1.45\cdot10^{6}$ & $3.87\cdot10^{4}$ & 0.638 & 0.342 \\
m20a   &  20  & $9.40\cdot10^{2}$ &  $2\cdot10^{-2}$ & RSG & $7.50\cdot10^{6}$ & $9.53\cdot10^{5}$ & $2.29\cdot10^{4}$ & 0.609 & 0.371 \\
m25a   &  25  & $1.19\cdot10^{3}$ &  $2\cdot10^{-2}$ & RSG & $5.97\cdot10^{6}$ & $6.99\cdot10^{5}$ & $1.67\cdot10^{4}$ & 0.593 & 0.387 \\
m13b   &  13  & $4.08\cdot10^{2}$ &  $1\cdot10^{-3}$ &RSG & $1.47\cdot10^{7}$ & $1.80\cdot10^{6}$ & $6.03\cdot10^{5}$ & 0.708 & 0.291 \\
m15b   &  15  & $1.01\cdot10^{2}$ &  $1\cdot10^{-3}$ & BSG & $1.22\cdot10^{7}$ & $1.41\cdot10^{6}$ & $4.17\cdot10^{5}$ & 0.769 & 0.230 \\
m20b   &  20  & $4.60\cdot10^{1}$ &  $1\cdot10^{-3}$ & BSG & $8.84\cdot10^{6}$ & $8.71\cdot10^{5}$ & $2.47\cdot10^{4}$ & 0.769 & 0.230 \\
m25b   &  25  & $5.26\cdot10^{1}$ &  $1\cdot10^{-3}$ & BSG & $7.09\cdot10^{6}$ & $6.69\cdot10^{5}$ & $1.80\cdot10^{4}$ & 0.769 & 0.230 \\
m15z   &  15  & $2.48\cdot10^{1}$ &  $0.0      $     & BSG& $1.05\cdot10^{7}$ & $9.27\cdot10^{5}$ & $5.44\cdot10^{4}$ & 0.770 & 0.230 \\
m20z   &  20  & $2.31\cdot10^{1}$ &  $0.0      $     & BSG& $7.74\cdot10^{6}$ & $5.76\cdot10^{5}$ & $2.39\cdot10^{4}$ & 0.770 & 0.230 \\
m25z   &  25  & $3.67\cdot10^{1}$ &  $0.0      $     & BSG & $6.72\cdot10^{6}$ & $4.96\cdot10^{5}$ & $1.59\cdot10^{4}$ & 0.770 & 0.230 \\
\hline
\end{tabular}                                                                               
\end{minipage}
\end{table*}

\begin{figure}
\includegraphics[width=80mm]{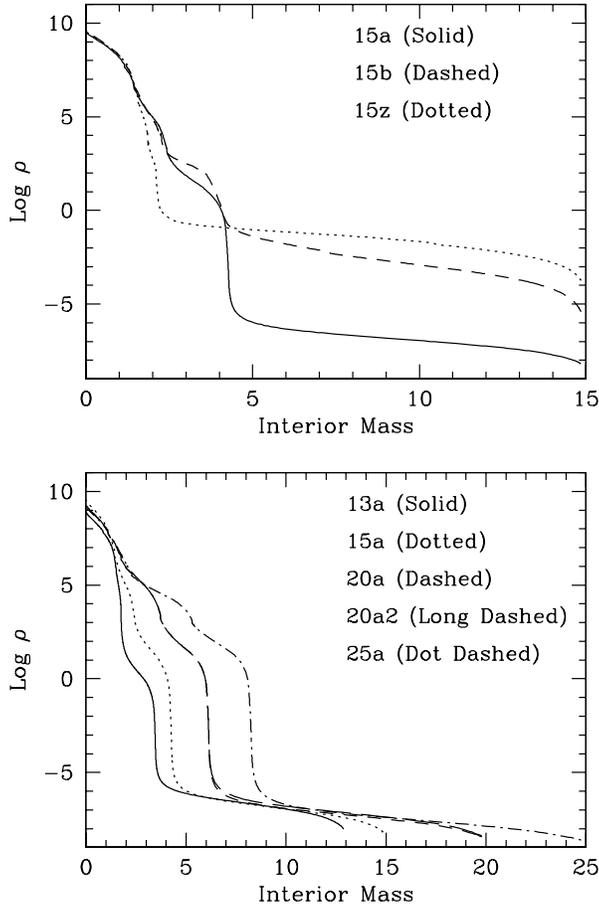}
\caption {Stellar density structure as a function of mass for different metallicities 
(upper panel) and initial masses (lower panel). Note the similarity between the density
  profiles of the two BSGs, 15b and 15z (upper panel). To test the convergence of the resolution in our models,
we have increased the number of depth points by a factor of 10 in model 20a, the resulting
structure, 20a2 (see lower panel), is almost identical.}
\end{figure}

Figure 2 shows the chemical composition  at the onset of the iron core collapse,
 for some of the computed stellar models. Figure 3 illustrates the influence  of
 both metallicity (upper  panel) and initial  mass (lower panel)  on the density
 profiles. As it is well known, the  smaller the total mass the less compact  is
 the star. By comparing Figure 2 and 3 it appears that the last (most  external)
 sudden drop in the density profiles corresponds to the transition to the H-rich
 envelope.  From  this  point  to  the  surface  the  density-mass  relation  is
 essentially independent  on the  initial mass,  while, on  the contrary,  it is
 significantly affected by a metallicity  variation. As it will be  discussed in
 more details in the following sections, such a correlation between the  density
 of the H rich envelope with the mass and the metallicity of the stellar  progenitor
 has  a significant influence on the  features of the various light curves,   in
 particular on the plateau phase (namely the first 100-150 days). In fact during
 this  phase,  the H  recombination  front, that  provides  the nearly  constant
 luminosity, moves inward (in mass) through the whole envelope. 

\begin{figure}
\includegraphics[width=80mm]{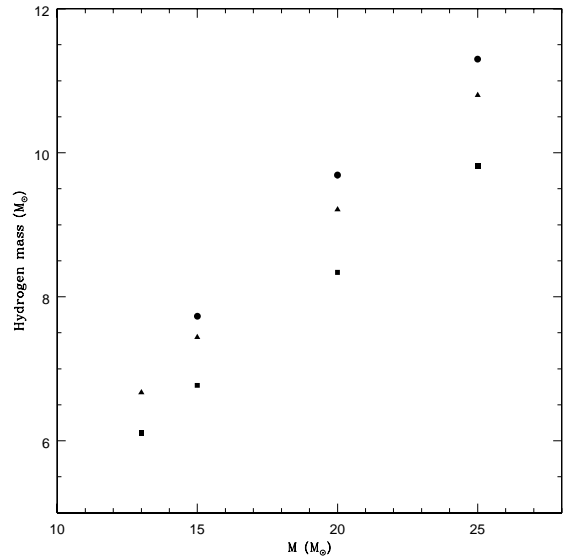}
\caption {Final  hydrogen mass as a function of total mass  for  
  different metallicities:  Z=0.02 (squares), Z=0.001 (triangles) and 
 Z=0  (circles)}
\end{figure}

Another  important quantity  that characterizes  the light  curve is  the  total
 amount of H present in the  envelope. In principle, it depends on  the original
 chemical composition of the  star, on the initial  mass, on the mass  loss rate
 and on the efficiency  of the various dredge  up episodes occurring during  the
 progenitor life. In Figure 4 we have reported the final H mass as a function of
 the total mass for  the 3 different metallicities.  Note that the final  H mass
 linearly increases as the stellar mass increases and the metallicity decreases.
 We find that  the following relation  nicely reproduces the  results of our stellar
 evolution calculations:

$\rm M_H(M_{\odot})$=2.58+0.338 $\cdot M(M_{\odot})$- 50.3 $\cdot Z$

where $\rm M_H$,  $\rm M$ and  $\rm Z$ are  the total H  mass, the total stellar
 mass and the metallicity, respectively. Since our models were obtained  without
 mass loss, this relation provides an upper limit for the final amount of H.     

\begin{figure}
\includegraphics[width=80mm]{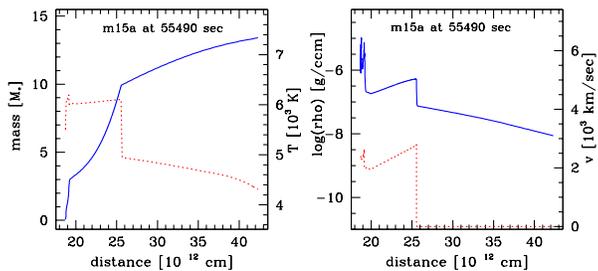}
\caption {
Structure of the 15 $\rm M_\odot$, Z=0.02 model  with a final kinetic energy of $10^{51}$ erg (model m15a)  
15.4 hours after the explosion.
Integrated mass M(r) (solid) and  T(r) (dotted) are given in the
left plot, $\rm \rho(r)$ (solid) and v(r) (dotted) are given in the right plot. 
}
\end{figure}

\subsection{Explosion Models}

Based on  the evolutionary  models previously  described, we  have explored  the
 sensitivity  of  different characteristics  of  the light  curve  on progenitor
 properties  (mass and  metallicity) and  explosion  energy. 
  The explosion is triggered by depositing a   given
 amount of energy at a mass coordinate close to the edge of the iron core of the
 presupernova model, i.e. at about 1.4 $\rm M_\odot$ that, in turn,  corresponds
 to about 1000 km in all the models. The injected energy is properly adjusted in
 order to provide the  desired final kinetic energy.  In all models except  one,
 the final kinetic energy is  10$^{51}$  erg,  the exception is model m15a;  for
 this  model  we  consider  also  the  case  with  a  final  kinetic  energy  of
 2$\cdot10^{51}$  erg  (hereafter  model  m15a2).  For  test  calculations,  the
 explosion  energy has  been deposited  both as  thermal or  kinetic energy  but
 little difference has been found in the explosion models. 
 
\begin{figure}
\includegraphics[width=80mm]{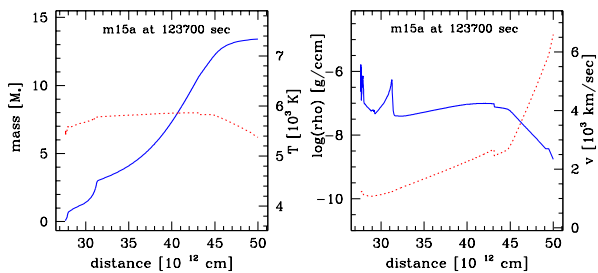}
\caption {
 Same as Fig. 5 but 34.3 hours after the explosion, just after shock
breakout.
}
\end{figure}

\begin{figure}
\includegraphics[width=80mm]{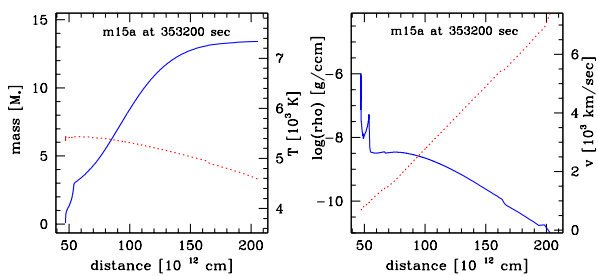}
\caption {
 Same as Fig. 5  but 98 hours after the explosion when the expansion of
the envelope is almost homologous.
}
\end{figure}

\begin{figure}
\includegraphics[width=80mm]{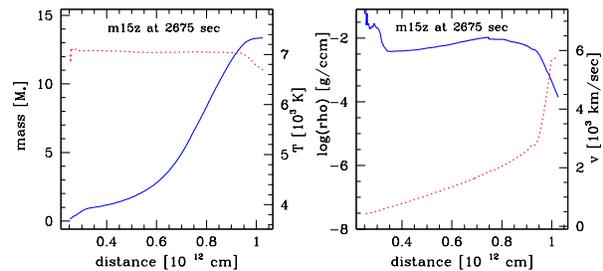}
\caption {
 Same as Fig. 5  but for the  model m15z, Z=0.0 
 at about 44 minute  after the explosion just after shock breakout.
}
\end{figure}

 In general, at a given time after the explosion has been  triggered, the
 velocity of the more internal zones of the exploding envelope drops down
 to the escape velocity. As a consequence, these zone will eventually and naturally 
 fall back onto the compact remnant. The final mass location between
 the ejecta and the remnant is defined as the mass cut. 
 The  fallback of  material on  the central  neutron star   remained
 smaller than  $10^{-2}$  $\rm M_\odot$  except  in the   more massive  models. A
 significant fallback of 0.1  and 0.47 $\rm M_\odot$  was obtained for m20a  and
 m25a, respectively. We note that  the amount of fallback depends  sensitively on
 the explosion energy.  In test calculations  for m25a with  twice the explosion
 energy, the fallback was reduced by about a factor of 5. In all cases, but  for
 25 M$_\odot$, more  than a tenth  of a solar  mass of $\rm  ^{56}Ni$ would be
 ejected which is  in excess  of the  typical amount
 obtained from the observed luminosity of the light curve tail. For this reason
 we introduced an {\it artificial} mass cut (more external than the actual mass cut)
 in the  explosion
 models  at about  one day  after the  explosion to  limit the  $\rm ^{56}Ni   $
 production accordingly. In this work  we
 adopt the typical value of 0.07 $\rm M_\odot$  (e.g. SN1987A) for  the
 $\rm ^{56}Ni$ mass,  although some SN II-P  are known to have more than a tenth
 (e.g. SN1992am, 0.3 $\rm  M_\odot$, Schmidt  et al. 1994;  1986I, 1991G, 1992H,
 Hamuy 2002). 

\begin{figure*}
\includegraphics[width=60mm,angle=270]{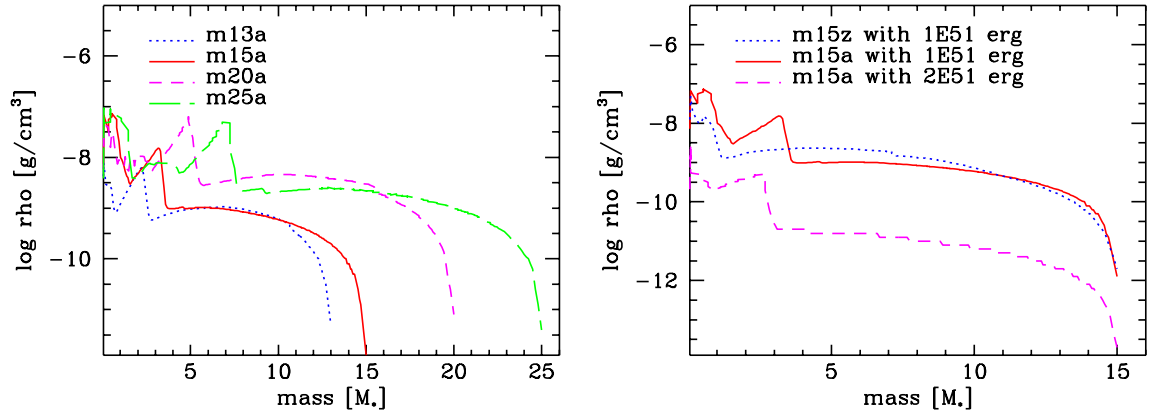}
\caption {
Density profiles at day 5  for models with various masses (left panel), metallicities
end kinetic energies (right panel).}
\end{figure*}
                                                                                                            
\begin{figure*}
\includegraphics[width=60mm,angle=270]{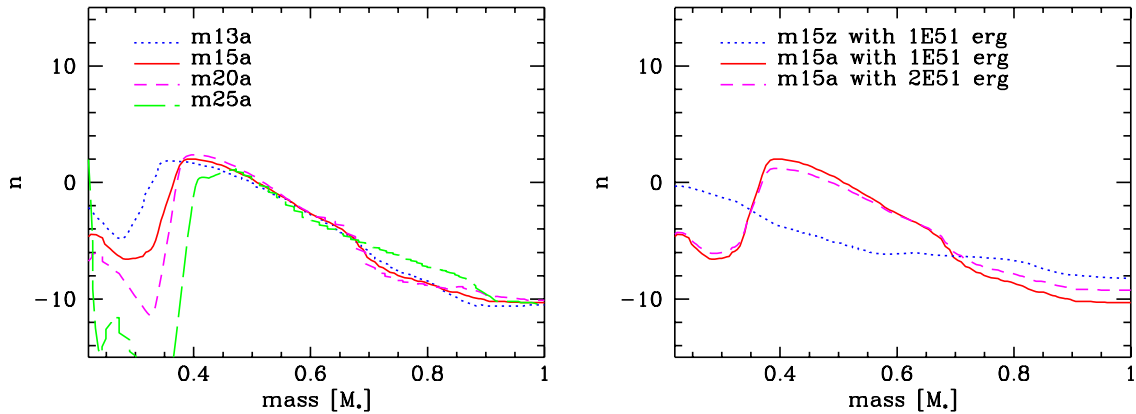}
\caption {
Same as Fig. 9 but for the density gradients $n$ ($\rho \propto  r^{-n}$). In the x axis
the interior mass is normalized to $\rm 10~M_\odot$.
}
\end{figure*}

 A further restriction of our models  is related to the                                                     
 discretization in mass which is of the order of                                               
 $\rm 10^{-3}~ M_\odot$.                                                                                           
 A proper resolution of the photosphere during the first few days of                                        
the explosion requires a discretization of about $\rm 10^{-5}~ M_\odot$                                          
 (e.g.                                                         
M\"uller \& H\"oflich, 1991).
 Consequently, details of the shock breakout are beyond the scope of                                        
this study because the photosphere is not well resolved at the early                                        
phases.                                                                                                     
                                                                                                            
\begin{figure}
\includegraphics[width=120mm,angle=270]{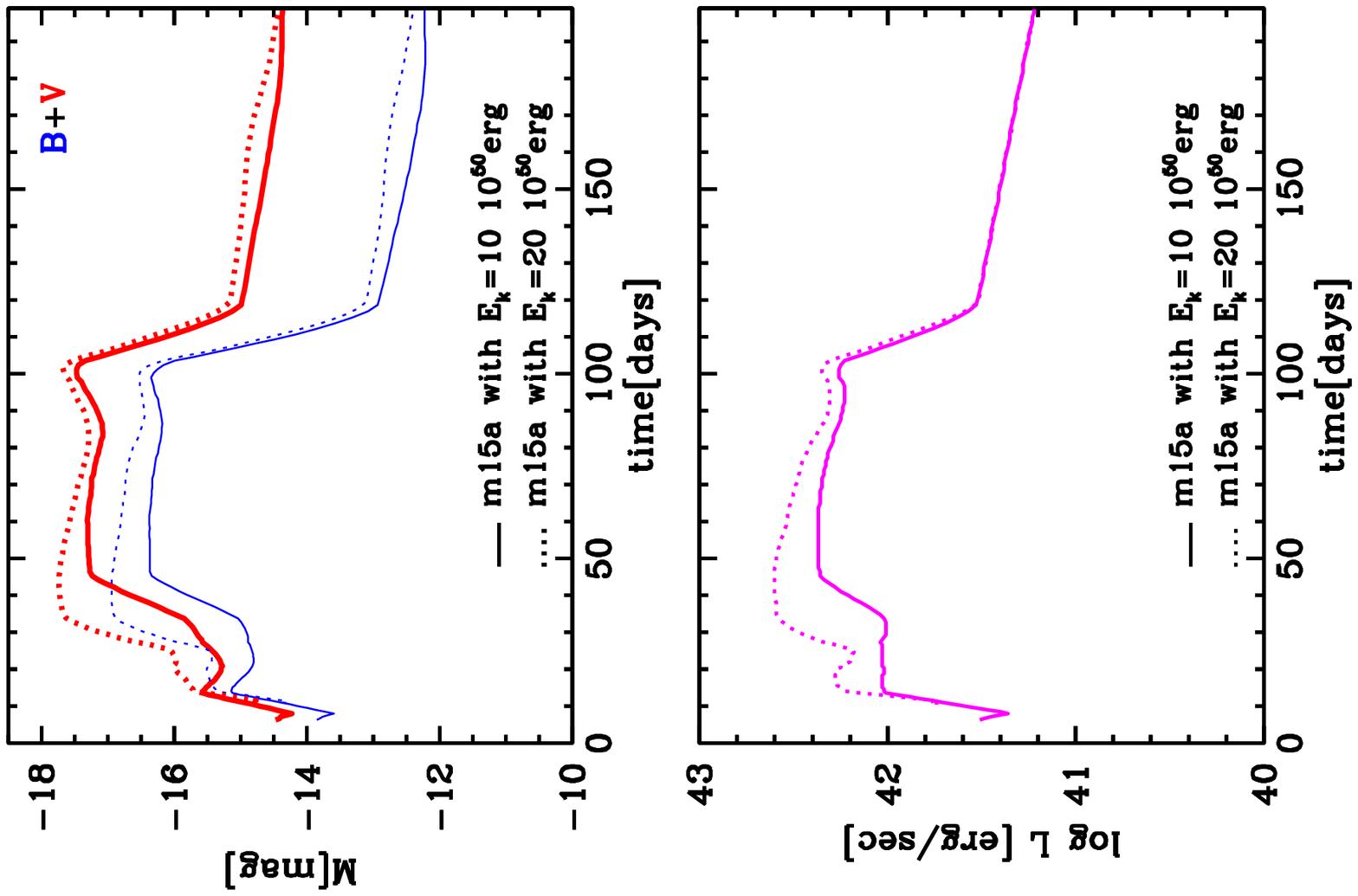}
\caption {  
Light curves for the 15 $\rm M_\odot$, Z=0.02 model with two  final
 kinetic energies, $10^{51} erg$ and 2$\cdot 10^{51} erg$, 
 models m15a and m15a2 in Table 2, respectively.
The monochromatic LCs in    V  and B (upper panel) are given by the thick and thin lines, respectively. The luminosity is shown in the lower panel.  
}
\end{figure}

 In Table 2, the basic parameters and some of the derived quantities                                        
are given for the explosion models and light curves. Let's concentrate here in 
 the first 4 columns:  identification of the model, as in Table 1  
  (column 1); 
  final  kinetic energy after the explosion  (column 2); time of the shock breakout in 
 seconds (column 3) and the corresponding temperature
 $T_{shock}$ at the photosphere (column 4).

\begin{table*}
\begin{minipage}{126mm}
\caption{Explosion models and Light Curve properties.}
\begin{tabular}{lclclcc}
\hline
Model &
  $E_{kin}$ [$10^{50}$ erg] &
$t_{shock}$  [s]  &
log( $T_{shock}$) &
 $V_{max}$ & 
$\delta t_{plateau}$ [days] &
 $\delta m_{bump}$ [mag] \\
\hline
 m13a & 10  & 1.33$\cdot10^{5}$    & 5.5 & -17.49 & 45 & 0.15  \\
 m15a &  10  & 1.21$\cdot10^{5}$ & 5.4 &  -17.47 & 66 & 0.22    \\
 m20a & 10  & 2.10$\cdot10^{5}$ &  5.3  &   -17.44 & 73 &  0.29 \\
 m25a & 10  & 3.10$\cdot10^{5}$ & 5.2   &   -17.42 & 78 & 0.39 \\
 m15a2  & 20  & 8.47$\cdot10^{4}$ & 5.6 &  -17.74  & 79 & 0.26 \\
 m15b & 10  & 9.23$\cdot10^{3}$ & 5.9  &   -16.43 & dna &dna \\
 m15z & 10  & 2.62$\cdot10^{3}$ & 6.7  &   -16.20 & dna &dna \\
\hline           
\end{tabular}
\end{minipage}
\end{table*}

 In Figs.  5 to 7, a typical evolution of the exploding star is given
at the example of model m15a. Initially, the shock front                                                  
propagates  outward and deposits energy in form of thermal energy.                                         
Weaker, reversed fronts are created at the chemical boundaries.                                             
For m15a, after about 1.5 days, the shock front reaches the outer stellar layers.                                     
 The shock front is accelerated because of the steep density profiles                                       
at the surface layers of the star                                                                           
(Fig. 3), and produces a rapidly
expanding outward layer (see Figure 6, right panel).                                                                         
 During the following time, most of the thermal                                                             
energy is used to overcome its potential, do expansion work,                                                   
 and accelerate the expanding                                                                            
envelope.   After about 3-4 sound crossing times of the progenitor,                                         
the expansion of the envelope is almost homologous, i.e. $v \propto r$,                                     
ending the phase dominated by hydrodynamics (see Fig. 7, right panel). 
 Subsequent energy release by radioactive decays causes only minor
modifications of the density profile.
 We note that, in reality, some interaction with the surrounding                                            
medium may become important.                                                                                
                                                                                                            
\begin{figure}
\includegraphics[width=80mm]{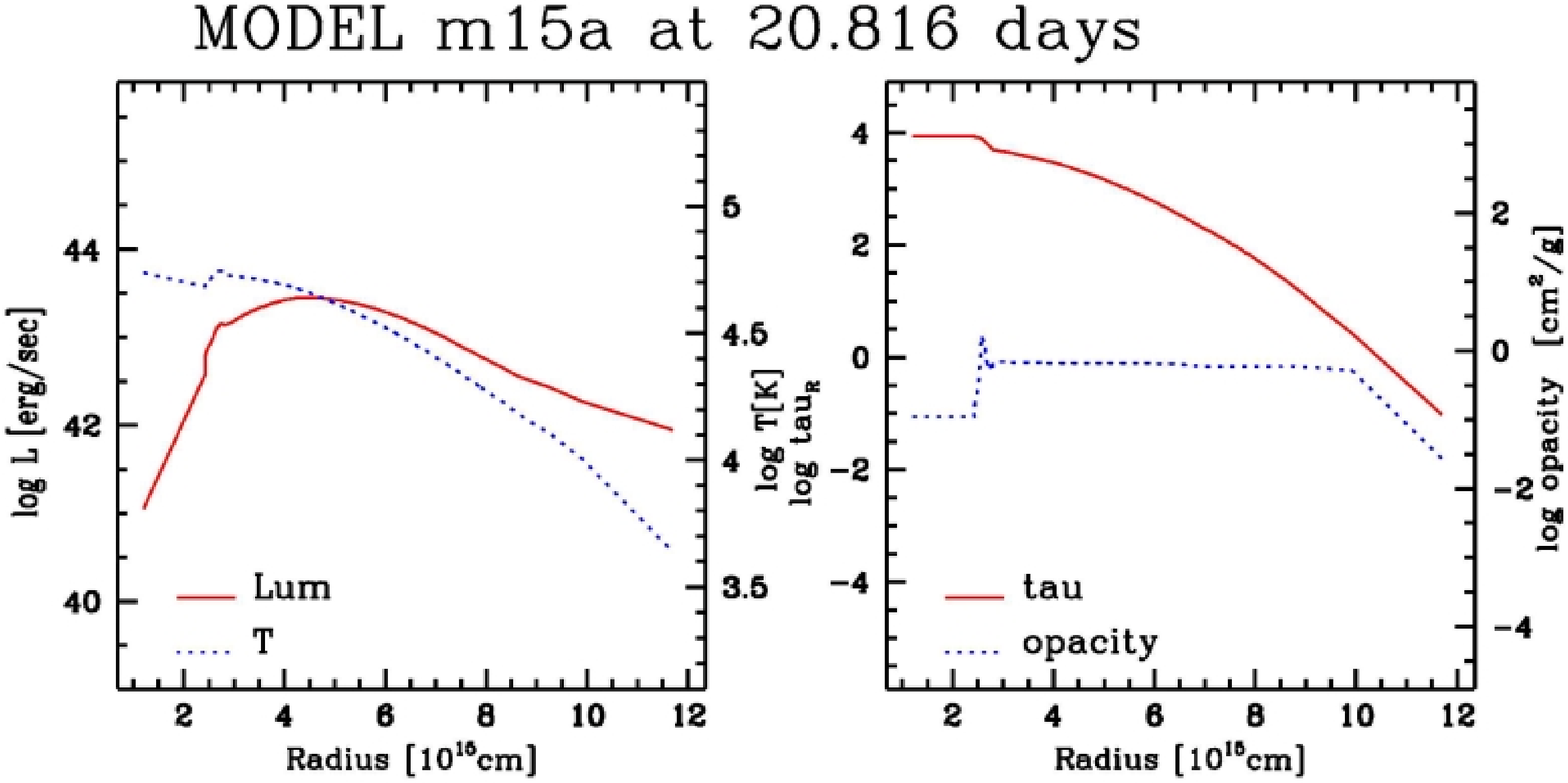}
\caption { 
Luminosity, temperature, opacity and optical depth for the 15 $\rm M_\odot$, Z=0.02  model (m15a)   
before the recombination phase, at day 20.
}
\end{figure}

\begin{figure}
\includegraphics[width=80mm]{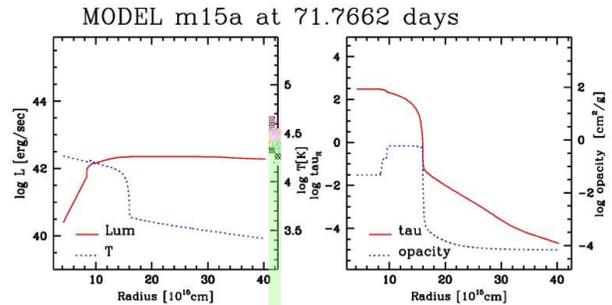}
\caption {
Same as Fig. 12 but 
during the recombination phase, at day 71.
}
\end{figure}

 All models show a behavior very similar to our example with some                                          
quantitative differences. For example, in case of the explosion of a                                        
compact, blue supergiant, the shock front reaches the surface already after about one hour             
 and the breakout temperatures are significantly larger, and                                        
subsequently, adiabatic cooling is increased drastically (see Table 2 and 
 Fig. 8).

The times till the shock breakout $t_{shock}$ (Table 2) are                                                           
consistent with the analytical approximation by Shigeyama et al. (1988)                                     
who found that  $t_{shock}$ scales with the  stellar radius,                                                
and with the square of the mass and  explosion energy.                                                      
  In our models, typical times are of the order of days for RSGs (1 to 3 days) and hours for BSGs 
 (1 to 3 hours).

 It is worth noting that the peak  temperatures for red supergiants                                         
are rather low whereas blue supergiants can reach peak temperatures                                         
in excess of a million degrees (see Table 2). The possible consequences for the                                           
contribution of the production of high energy photons at large redshifts                                    
shall be noted and its implications for environment will be discussed                                       
in the Conclusions.                                                                                                     
                                                                                                            
Density structures and density gradients  are given in Figures 9 and 10, respectively.
As can be expected from the discussion of the stellar profiles (Fig. 3),
the final density structures are rather similar for the red                                                 
supergiants. For the blue supergiants, the density profiles are                                              
significant steeper in the inner layers of the hydrogen rich envelope
which, as we will see, has strong effects on the light curve.

\subsection{Light Curves}
 In this section, we will discuss the phase                                   
when the properties of the  envelope are mainly determined by                 
free expansion and radiative processes and                                       
the energy release is governed by stored energy, thermal energy,                        
recombination processes and radioactive decay.

 In general, an early maximum is seen which is produced by the                
release of the stored thermal energy, followed by a plateau phase             
due to the recombination of H and, finally, a long tail due to the energy          
release by radioactive decay of $\rm ^{56}Co$ (e.g. Fig. 11).
                      
In the last 3 columns of Table 2 we show some properties of the light curves: the maximum visual  magnitude (column 5), 
 the length of the plateau phase, in days (column 6) and the size of the bump, 
 in magnitudes, occurring at the end of the plateau phase (column 7). The length of the 
 plateau is defined  by the times when $M_V$ becomes larger and smaller  than $M_V (max) + 0.6^m$.
 In the table "dna" means do not apply.
                                                                                
 Firstly, we consider the evolution of the structure for our          
reference model m15a, and the corresponding light curves                   
(Figs. 11-13). The initial flash in the light curve is due to the energy deposition
at shock breakout. Its duration is of the order                 
of the sum of cooling time, the light crossing time of the stellar            
radius and the shock traveling time through the photosphere.                  
 For red supergiants, it is of the order of a few hours, for our blue          
supergiants about 10 minutes.                                                 
 For about 3 weeks after the explosion, the photospheric temperature          
 is sufficient high to maintain ionization up to the outer layers                 
(Fig. 12). The opacity is dominated by Thompson scattering, bound-free and
free-free processes in the optical and IR,  and by line                       
blocking in the UV, resulting in very high optical depths of the               
envelope. Consequently, the expansion of the photosphere is strongly          
coupled to the expansion of the material.                            
 The diffusion time scales for photons, $t_{diff}$,  are given                  
by    $$ t_{diff}(r) \approx \tau(r)^2 \cdot r/c $$                                
where  $\tau$ is the optical depth,  $r$ the radius and $c$ the speed of          
light. In our example at day 20, the diffusion time scales exceed             
the expansion time scales in all layers up to about an optical depth          
of 10. As a consequence, the luminosity as a function of depth is not constant but it
is increasing  inward due to  the stored energy released by the shock front and                      
the receding (in mass) photosphere (Fig. 12). For the same reason, models with
low explosion energies show a brightening between 30 to 50 days after the explosion.

 After the initial rise, the reference model shows an {\it early plateau} not seen
in observations. It is caused by the 1D nature of the model or, more precisely, due to
the assumption that Ni is not mixed. For models with moderate explosion
energies, the diffusion time scales for energy stored by radioactive decays
is longer than the expansion time scale. As a consequence, the luminosity is
solely provided by the thermal energy stored during the explosion because
the contribution by radioactive decays is delayed.
 This effect is well known from models for SN1987A where mixing had to be
assumed to avoid this artifact and to obtain good fits to the light curves observed
(e.g. Woosley et al. 1988). Note that we do not see this early plateau in any
observation. This may be a hint that the explosion mechanism is intrinsically aspherical
 and  that mixing of the central layers is common in core collapse supernovae
 (see introduction).

 With time, the photospheric temperature drops, and recombination of H         
sets in. Due to the strong drop in opacity with decreasing ionization,        
 the position of the photosphere becomes  almost stationary                   
(Figs. 14).  In Fig. 13, the structure of m15a
is given at day 70 which is typical for the recombination phase.              
Diffusion time scales for the envelope become comparable to the               
expansion time scales. The luminosity is governed by the                      
release of recombination energy which is deposited just below the              
photosphere.                                                                  
 Typically, the ionization degree at the                        
photosphere has dropped to 1 \%. Therefore, the luminosity as a function      
of radius increases outward up to the photospheric radius and stays constant          
for the outer layers.                                                         
 The rate of the energy release depends on the recombination rate             
(in mass) and this depends on the mass flow through the photosphere, i.e.          
on the density slope $n$ ($\rho \propto  r^{-n}$). 
Since the density of the envelope is rather flat and slowly changing
a self-regulating mechanism between                                             
energy release and heating of the photospheric region leads to an
almost constant luminosity of the light curve.
An increase in
the energy release causes a heating of the photosphere and, thus, a higher degree of ionization and
a larger opacity which, in term, reduces the luminosity and vice versa.

 Eventually, the recombination front reaches the hydrogen-poor            
layers, and enters the He-core.
 At this point, the photosphere recedes very fast causing
an energy release and a small bump in the light curve at the end of the
 recombination/plateau phase (see Fig. 11 and below).
 Thereafter, the energy is purely determined
by the instantaneous energy input                                               
 due to radioactive decays, mainly by $\rm ^{56} Co$,                                
up to a few hundred days after the explosion, then diffusion time scales become    
negligible. Except for models with extensive mixing of          
$\rm ^{56}Ni$, the envelope above the radioactive elements remains               
optically thick for $\gamma$-rays and the high energy photons thermalize            
within the envelope almost completely.               
          
 At the onset of the recombination phase, B-V increases rapidly due to the drop in temperature and the
increasingly strong line blocking. Subsequently, during the recombination phase, B-V changes slowly
 from  about 1.2 to 1.5 magnitudes because the conditions remain similar at the photosphere.
 At the end of the recombination phase, again, B-V drops rapidly to
about 2 magnitudes due to the strong line blocking in
the B band, and due to a further decrease in temperature.

At the end of the plateau phase,  the reference model shows brief period of increasing
brightness in B and V, and bluer colors. This
is caused by the rapid change in the He abundance at the photosphere which results in an
increased recombination temperature and a more rapidly decreasing photosphere
 and, thus, a temporary
increase in the release of stored energy. These effects should be strong in SNe with a low explosion 
energy.  Both the late increase brightness and the decrease in B-V have yet to be observed. 
The lack of evidence for these
effects may be regarded as further hint that, in reality, the inner layers of SNe are strongly mixed
during the explosion.
                                                                              
\begin{figure*}  
\includegraphics[width=60mm,angle=270]{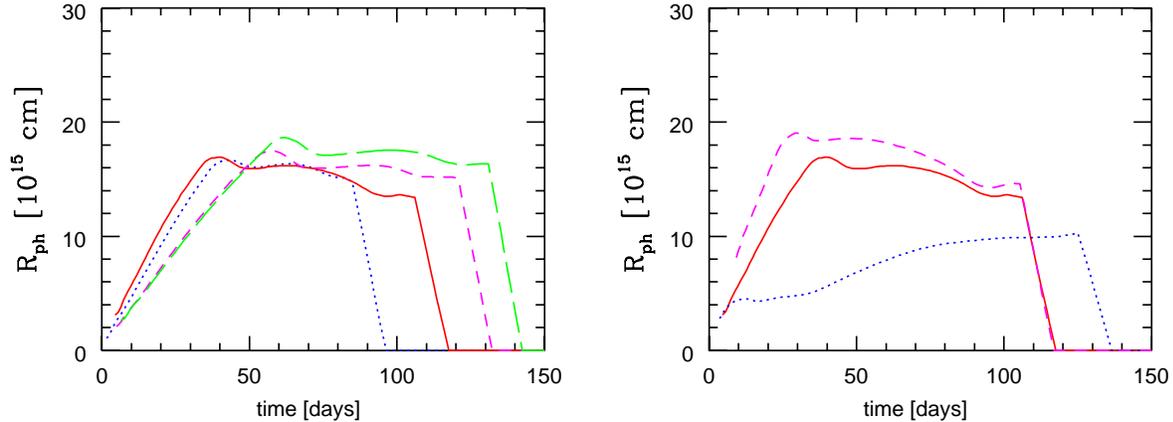}
\caption {
Photospheric radius as a function of time for models with the same 
 metallicity, Z=0.02, but with different 
 initial masses: 13 $\rm M_\odot$ (dotted), 15 $\rm M_\odot$ (solid), 20 $\rm M_\odot$ (dashed) and 25 
 $\rm M_\odot$ (long-dashed) on the left panel and 
 for models with the same mass, 15 $M_\odot$, but different metallicities and/or kinetic energies on the right panel: 
 Z=0.02  and 1 foe (solid), Z=0.02 and 2 foe (dashed) and Z=0 and 1 foe (dotted) on the right panel.
}
\end{figure*}

The influence of the kinetic energy on the light curves                       
is shown in Fig. 11.
 Overall, the shape of the light curve shows only some quantitative           
changes.                                                                      
 Increasing the kinetic energy by a factor of 2 results on  a faster rise of                                    
the light curve, an increased luminosity at the plateau ($\approx$ 0.27$^m$ in 
 the V band), and a slightly      
bluer color during the plateau phase.                                         
 The overall similarities at the plateau are due to the similar density structures, as seen in the last section        
(see Fig. 9 and 10).
The increase of $E_{kin}$ results in a faster expansion rate of the            
material by about 40 \%, and an increased energy deposition due to the        
shock front.                                                                  
 Due to the faster geometrical dilution, the stored                           
thermal energy is released faster, causing an increase in the early             
 luminosity. During the early               
recombination phase, thermal energy still contributes to the flux.            
The increased mass flux through the  photosphere increases                   
the luminosity by about 30 to 40 \% according to the increase                 
in the expansion velocity. The higher flux results in a slightly bluer        
color of the more energetic model.                                           
 In both models, we assume the same                                           
$^{56}$Ni ejection and, consequently, the light curves become very similar   
after day 130 and the luminosity of the light curve tail are identical. However,          
the lower density requires a slightly higher color for the more energetic model to              
maintain the same integrated emissivity.

In Figure 15, we show the light curves for our set of models with
 Z=0.02 and different initial masses: 13, 15, 20 and 25 $\rm M_\odot$, all of which explode as 
 RSGs. The maximum brightness and the overall shape of the light curves remain very similar, 
 showing a long plateau phase, in excess of 50 to 60 days and extending up to 80-130 days from 
 the explosion time. We identify this group as an homogeneous subclass among Type II-P, i.e., 
  {\it  extreme} SN~II-P. The mean brightness in V during the plateau ($\approx$ -17.4) is 
  rather insensitive to the mass of the progenitor ($\Delta M_V \le$ 0.07 mag), and to
 the explosion energy if changed within a factor of 2. \footnote{The explosion energy may be
constrained by spectral observations}. This suggest that
 this subclass, {\it  extreme} SN~II-P, may be used as quasi-standard candles with few free
parameters, namely the Ni mass.

 The similarity of the light curve shapes  can be understood as a consequence of the                         
similarity of the density slopes $n$ (Fig. 10) that results
in comparable energy production rates due to recombination.                   
 The main difference among the light curve is the length of the plateau phase because               
of the increase in the total hydrogen mass that is available as a                 
reservoir for storing ionization energy (see Table 2).                      

 Compared to Litvinova \& Nad\"ezhin (1983, 1985), our RSG models show
a similar correlation between the brightness during the mid plateau stage and the
kinetic energy but a much weaker relation to the envelope mass. We have used  the relations 
 (4) and (5) from Litvinova \& Nad\"ezhin (1985) to compute the mean V magnitude during  
 the plateau and its duration in function of the explosion energy,  R of the progenitor 
   and the envelope mass. We obtain, for our mass range (envelope masses between 
 11.6 M$_\odot$ and 23.6 M$_\odot$) and in contrast with our models, that the mean V 
 magnitude increases with mass; a variation of 0.23$^m$ is obtained through the whole 
 mass interval to be compared with the variation of 0.07$^m$ resulting from our models. 
 The duration of the plateau obtained using  Litvinova \& Nad\"ezhin equation increases 
 with envelope mass as expected but, for all cases, is about 20 to 50 \% times longer than
 the duration resulting from our models.    
 These discrepancies can be understood as a consequence of the progenitor structure (see introduction).
Litvinova \& Nad\"ezhin (1983, 1985) use parameterized density structures with the stellar radius as a free parameter
whereas our models are based on stellar evolution. In particular, Litvinova \& Nad\"ezhin (1985) change the
the envelope mass from 16 to 1 M$_\odot$  but assume  the same radius. As a consequence, the column
densities of the envelope differ by a factor of 16 and, consequently, the speed of the energy release.
 In contrast, our models m20a and m15a have an envelope mass of 18.6 and 13.6 $\rm M_\odot$ but the radius decreases  
from 970 to 670 $\rm R_\odot$ and the column densities in the envelope at a given radius are rather similar.

                                                                              
\begin{figure}
\includegraphics[width=120mm,angle=270]{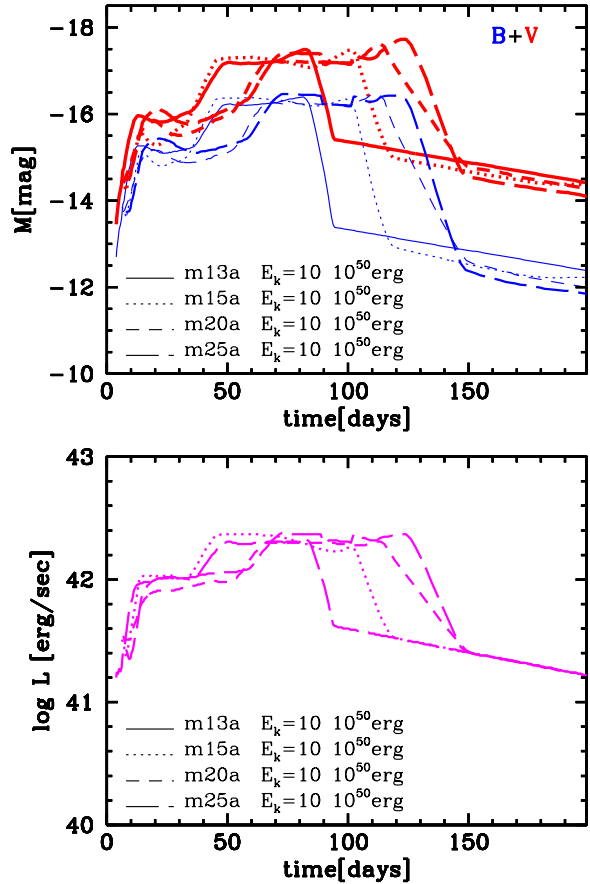}
\caption {Light curves for models with the same composition (Z=0.02) and 
 final kinetic energy ($10^{51}$ erg)  but different initial masses:   
13 $\rm M_\odot$ (m13a), 15 $\rm M_\odot$ (m15a),  20 $\rm M_\odot$ (m15a) and 25 $\rm M_\odot$
 (M25a). 
The monochromatic LCs in V and B are given by the thick and thin lines, respectively.
} 

\end{figure}
                                                                              
 For a blue supergiant, i.e. m15z, the density slopes are much steeper     
compared to our reference model m15a (Fig. 10). Starting from a more compact
envelope the cooling by adiabatic expansion increases, while 
the overall luminosity decreases by more than a magnitude. 
The maximum V magnitude at the plateau phase decreases by  1.3$^m$ (Table 2). 
In this case the light curves do not 
 show the long plateau phase (see Fig. 16) but a slowly
rising up to the maximum light. 
This is due to the steep density slope that causes a strongly increasing               
recombination rate as a function of time.

As a test to confirm this results, we have also computed the explosion and light curves for models with Z=0.0 and 
initial masses 20 and 25 $\rm M_\odot$, as expected from the density gradients, the light curves show the same properties as  m15z. 
Note also that line blocking in B and, in particular, in the UV, depends also on the metallicity, decreasing B-V for the low metallicity models.

 
\begin{figure}
\includegraphics[width=120mm,angle=270]{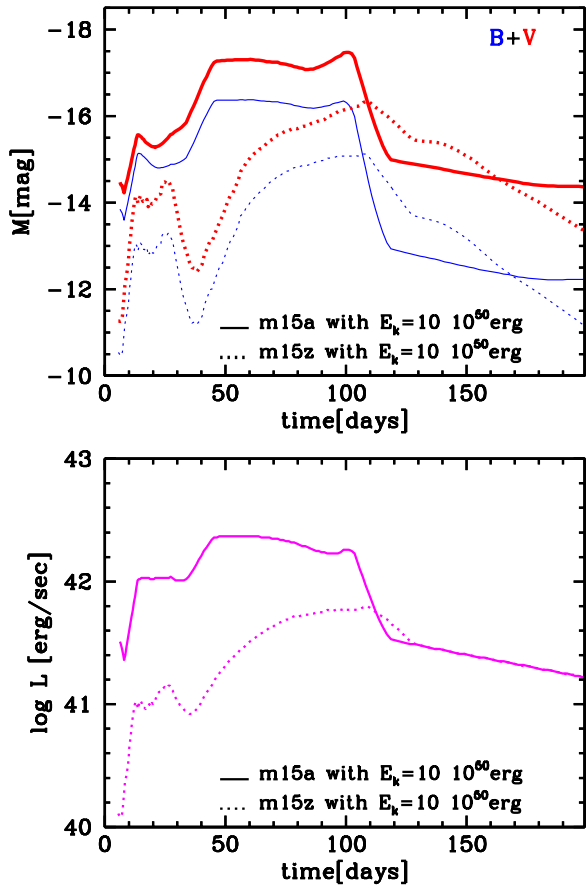}
\caption { 
Light curves for models with the same mass, 15 $\rm M_\odot$, and kinetic energy, $10^{51}$ erg, but different metallicities, Z=0.02 (m15a) and Z=0.0 (m15z). 
The monochromatic LCs in    V  and B  are given by the thick and thin lines, respectively.
}  
\end{figure}

\section{Conclusions}
We have studied theoretical  B, V and bolometric light curves for type II plateau supernovae. Based on a set of
 stellar evolution calculations, we have analyzed the sensitivity 
 of the light curves properties to  the mass and  metallicity  of the progenitor and, within a factor of 2,
 to the explosion energy.
 We linked the  pre-SN evolutionary models to SN models. Compared to previous studies, one big advantage
is that the initial  radius of the SN models is no longer an adjustable parameter but
a result of the physical model and a function of the stellar mass, changing significantly  the relation
between progenitor mass and light curves.
 The other   progress is that the hydro models are linked to a radiative transport code,
so the broad-band light curves are much more accurately computed.
A third  progress provided by this study  is the effect  of metallicity in the resulting
light curves.
 
We find the following main results within the parameter space considered:
 
For high metallicities, the stars explode  as RSG regardless of the initial mass, showing 
 a long plateau phase, longer than 50-60 days and extending up to 80-130 days after the explosion ({\it extreme} SNe~II-P).
 They can be understood as explosions of Red Supergiants that have undergone rather moderate
mass loss during the presupernova evolution.
 The V brightness during the plateau phase changes/declines by about 0.2 to 0.7$^m$.
The mean absolute brightness in V ($\approx -17.5^m$)
during the plateau phase is rather insensitive 
to the mass of the progenitor and the explosion energy  (within $\approx 0.3^m$).
 Note that line blocking in B and, in particular, in the  UV depends on the metallicity  causing a
somewhat larger spread.

The overall similarity of the LCs is caused by the similarity  of the density structures
of the red giant envelopes, and by the
 self-regulating propagation of the recombination front that
determines the brightness during the plateau phase.
 For  explosion energies less than $10^{51}$ erg, the self regulating mechanism between
photospheric radius and the location of the recombination front will break down because the mean
temperature of the envelope will drop below the recombination temperature. As a consequence,
the absolute brightness during the plateau should drop fast with $E_{kin}$.

The metallicity plays a major role in determining the radius of the star
 because it affects the opacity that determines directly the super-adiabatic
  gradient that holds in the more external layers. The larger the opacity, the larger the 
 super-adiabatic gradient and hence the larger the radius of the star. This explains why for 
 low metallicities, stars explode as blue supergiants.
The steep density profile
results in a long lasting phase of increasing photospheric radius and brightness. The maximum brightness is
 lower by  about $1.5^m$ compared to the explosion of a RSG because of the increased expansion work for BSGs.
Qualitatively, this tendency is consistent with SN1987A.
However as  shown above, a 13 $\rm M_\odot$ star will explode as a RSG even for Z as low as a $10^{-3}$.

  The mass dependence of the final outcome has  two main consequences. Firstly, the discovery
probability for SNe II at high $z$ will decrease with the progenitor mass. 
 The supernovae statistics will be systematically biased, starting at $z \approx 1$.
  The consequences for the study of the chemical evolution and the element production at high red-shifts
(e.g. by NGST) shall be noted.
Secondly, even at high redshifts, some {\it extreme} SNe~II-P should be visible. Taking their unique
properties, they may prove to be the key for the use of SN for cosmology at high $z$ before
SNe~Ia occur.
 It is worth noting, that the
photospheric  temperatures at the
 shock breakout are higher than in the solar metallicity models. 
  As the cross section for H-photoionization decreases with increasing frequency,
 $\sigma_\nu \propto \nu^{-3}$, the ionized circumstellar region around these SNe would 
 be larger compared to the RSG case. The possible consequences for the re-ionization
in the early universe may be noted.
 
 Our results may be suggestive to
the use of a subclass of SNe~II, the {\it extreme} SNe~II-P, as quasi
standard candles.
 Although the  use of {\it extreme} SNe~II-P will not achieve the same accuracy as Type Ia Supernovae, there are some
distinct advantages: 1) due to their unique light curves and colors, no spectrum is required for
identification; 2) the requirements on the time coverage of the light curves are very moderate:
 three or four deep images with a sample rate of 50 to 60 days in the rest frame
  will allow their discovery, 
 identification and their use for cosmology - at some time, two color images should be taken to
deselect flare star and to get a  handle on the reddening;
  3) finally, there is no need to follow the light curves
after the plateau toward dimmer magnitudes. For the use of SNe~Ia, the requirement to obtain a
spectrum limits their use  to $\approx 24^m$ if 8m-class telescopes are employed.
For the {\it extreme} SN~II-P,  1) to 3) implies that the largest ground based telescopes 
 with IR detectors can  be used as search instruments which
pushes the limit to about $27 $ to $28^m$. Therefore, {\it extreme} SNe~II-P may be used up to $z \approx 3$
 using 8-meter class telescopes.
  SIRTIF may push the limit by another magnitude by long time exposures.
Our results may be interesting with respect to the use of supernovae
as distance indicators, and the supernovae statistics that may be
constructed based on future observations by upcoming instruments
such as NGST, SNAP and the like.  
 For moderate to high metallicities, {\it extreme} SNe~II-P may be used as standard
candles with an accuracy of about 30\% if, in addition, the color information is 
taken into account.
 Though not comparative
with SNe~Ia at  low redshifts, their use may provide a valuable tool 
to supplement SNe~Ia distances in our local universe.

 At moderate 
redshifts (e.g. $z$=3-3.5), intergalactic metals have been discovered in the Ly$\alpha$
 forest clouds showing metal abundance of 10$^{-2}$ to 10$^{-3}$ Z$_\odot$ 
  (Songaila 1997, Cowie and Songaila 1998 and  Ellison et al. 2000), so it will 
  still be possible to   
 find {\it extreme} SNe II-P as distance indicators.

We have also to stress the limits of our investigations which will require further studies.
For our purposes, we have explored  a limited range of parameters. Recent observations
show a wide range of explosion energies and Ni masses well beyond the 'classical' estimates
of 1 and 2 foe and 0.07 $M_\odot$ of Ni.

 Hamuy (2002) found  Ni masses between 0.0016 to 0.26 $\rm M_\odot$, and  
kinetic energies from $0.6\cdot 10^{51}$ to $5.\cdot 10^{51}$ for envelope masses between 14 to 56
$M_\odot$.  Hamuy's interpretation for the kinetic energy and progenitor  mass is based on 
extrapolation of the empirical relations by Litvinova \& Nad\"ezhin (1983, 1985) which have been 
obtained for envelope masses between 1 and 16 $\rm M_\odot$ and on parameterized structures
(see also Section 3.3). Thus, the masses and explosion energies may be very uncertain.
 The importance of the progenitor structure becomes also evident in a direct comparison between
estimated progenitor mass of SNe II-P based on LCs and non-detections of progenitors.
Based on SNe observations and  on hydrodynamical models (Litvinova and Nad\"ezhin, 1983, 1985), Hamuy obtained progenitor masses
of  43(+24/-14) $\rm M_\odot$ and 27(+14/-18) $\rm M_\odot$ for SN1999gi and SN1999em, respectively.
 From the length of the plateau phase for SN1999em,
 H\"oflich et al. (2000) found models with $\rm 15 M_\odot$ to be consistent with observations.
Based on non-detection of the progenitor on archive images and stellar evolution,
Smartt (2001,2002) found  upper mass limits of 9(+3/-2) $\rm M_\odot$ and 12$\pm$1 $\rm M_\odot$ for 
SN1999gi and SN1999em, respectively.
 However, despite the problems, Hamuy's  estimates of the
total amount of $^{56} Ni$ are hardly effected by the model assumptions.
 A significant change in Ni will change the absolute brightness because radioactive decay contributes
to the thermal reservoir feeding the light curves.

 As discussed in section 3.3, mixing of the central layers will change the early rise and the
late plateau phases. Such mixing must be expected from explosion models and should be included to
improve the accuracy of the models.

No mass loss  has been taken into account.
Therefore, these models must be considered  as extreme cases, limiting severely their use 
for the analysis of observations. In particular, strong mass loss of RSG will effect the
relation between the envelope mass and the stellar radius differently from the change of the
initial mass that determines the structure of the stellar core. Likely, this will increase
the spread in parameter space.
One further potential pitfall for SNe II is the anisotropic luminosity caused by aspherical explosions of core collapse SNe.
 In general, the light of core collapse supernovae is polarized by $\approx  0.5\pm 1 \%$
(e.g. Wang et al.  2001, Leonard et al. 2001, 2002). Polarization of this size corresponds to asymmetries in the envelope that
produce directional dependence in the    observed L of $\approx 0.3 ~ to ~0.6^m$   (H\"oflich, 1991).
 However, extended H rich envelopes tend to spherize the H-rich layers of the
  envelopes even if the explosions are assumed jet-like (H\"oflich, Khokhlov \& Wang, 2001). 
This tendency is consistent with recent observations for SN1999em.

In summary, the typical light curves and colors of our {\it extreme} SN~II-P models are, at first order, in agreement 
with observations but  a quantitative comparison and 
statistical analyzes with observations must be postponed till
complete data sets become available, and the model grid has been extended.
 If done, such comparison may be
used to get an insight of details of the physics, like  
fallback during the explosion, realistic predictions for the
nucleosynthesis and constrains for core collapse models.
 For the application to cosmology and to get a handle on the interstellar reddening,  the
color information must be considered and spectral information may be used to increase the accuracy.
 Taking current initiatives, e.g. CSP (Carnegie Supernova Program), 
LOTOSS (Lick Observatory and Tenagra Observatory Supernova Searches), NEAT (Near-Earth Asteroid Tracking), 
SDSS (Sloan Digital Sky Survey) and the upcoming SNAP and NGST missions,
we expect an increasing availability of high quality data.

 \subsection*{ACKNOWLEDGMENTS}
 
PAH would like to thank the people at the Observatories in Teramo
and Monteporzio (Italy) for the hospitality during his stay when  
this paper was born. ID thanks the  people at the Dept. of Astronomy (University of Texas) at Austin for their hospitality during her stay when this paper was finished and thanks Mario Hamuy for his careful reading of the manuscript.
This research was supported in part by  NASA Grant LSTA-98-022, by the Italian grant MURST-Cofin2000,
  by the Spanish grant AYA2000-1574, AYA2002-04094-C3-03 and
 by the Andalusian grant FQM-292.
 The calculations for the explosion and light curves were done on a cluster of workstations 
financed by the John W. Cox-Fund  of the Department of Astronomy at the 
University of Texas.


\end{document}